\documentclass[%
 reprint,
superscriptaddress,
 amsmath,amssymb,
prx,
floatfix,
]{revtex4-2}

\usepackage{graphicx}
\usepackage{dcolumn}
\usepackage{bm}
\usepackage{xcolor}

\usepackage{bbold}
\usepackage{xparse}
\usepackage{hyperref}
\usepackage{siunitx}
\usepackage{cleveref}
\usepackage{braket,bigints} 

\usepackage{bbm} 

\renewcommand\bra[1]{{\langle{#1}|}}
\makeatletter
\renewcommand\ket[1]{%
\@ifnextchar\bra{\k@t{#1}\!}{\k@t{#1}}%
}
\newcommand\k@t[1]{{|{#1}\rangle}}
\makeatother

\usepackage{color}
\definecolor{mygreen}{rgb}{0,0.5,0}
\definecolor{mygrey}{rgb}{0.5,0.5,0.5}
\definecolor{myred}{rgb}{0.75,0,0}
\definecolor{myblue}{rgb}{0,0,0.75}
\definecolor{mymagenta}{cmyk}{0,1,0,0.12}
\definecolor{mycyan}{cmyk}{1,0,0,0.12}
\definecolor{myorange}{rgb}{1.,0.5,0}
\definecolor{myviolet}{rgb}{0.6,0.15,0.6}
\definecolor{mybrown}{cmyk}{0,0.50,1,0.41}

\usepackage{ulem}

\usepackage{mathtools}

\begin{document}

\preprint{APS/123-QED}

\title{Single-beam double-pass miniaturized atomic magnetometer for bio-magnetic imaging systems}

\newcommand{\State}{State Key Laboratory of Precision Measurement Technology and Instruments, Tsinghua University, Beijing, 100084, China}
\newcommand{\Dep}{Department of Precision Instruments, Tsinghua University, Beijing, 100084, China}

\author{Xiaojie Li}
\affiliation{\State}
\affiliation{\Dep}
\author{Zhihao Guo}
\affiliation{\State}
\affiliation{\Dep}
\author{Rui Yang}
\affiliation{\State}
\affiliation{\Dep}
\author{Yanying Feng}
\email{yyfeng@tsinghua.edu.cn}
\affiliation{\State}
\affiliation{\Dep}




\date{\today}

\begin{abstract}

Miniaturized atomic magnetometers, particularly spin-exchange relaxation-free atomic magnetometers, have been emerging in clinical imaging applications such magnetocardiography and magnetoencephalography. Miniaturization, portability, and low cost are primary development targets for bio-magnetic imaging technologies, as well as high sensitivity and spatial and time resolution. In this paper, we propose a low-cost solution for a bio-magnetic imaging system based on atomic magnetometers, in which one laser source is used for a multi-channel atomic magnetometer sensor array. A novel design is demonstrated for a miniaturized spin-exchange relaxation-free atomic magnetometer, consisting of a single-beam double-pass configuration based on an optical fiber circulator. The effects of temperature and laser power on the zero-field magnetic resonance line-width are characterized, and the experimental results show that the present design achieves better performance, than a traditional single-beam single-pass configuration. The noise power spectrum shows that the closed-loop miniaturized atomic magnetometer reaches a sensitivity of approximately 120 fT/Hz$^{1/2}$ at a bandwidth of 10 Hz. This design is especially suitable for atomic magnetometers operating in arrays as a basic building element for low-cost bio-magnetic imaging systems.

\end{abstract}

\maketitle


\section{Introduction}
\label{sec:Introduction}

Magnetic fields produced by biological organisms have recently received much attention owing to the valuable information contained in these fields regarding underlying physiological processes and their pathologies\cite{D. Cohen1968, H. Xia2006, G. Bison2009}. Over the past 40 years, these weak bio-magnetic fields have generally been detected by superconducting quantum interference devices (SQUIDs)\cite{M. Hamalainen1993, H. Weinstock1996, R. L. Fagaly2006}, which are commercially available but are still expensive in terms of both capital and operational costs.

Rapid progress has been made in the field of atomic magnetometers (AMs), which have recently achieved sensitivities comparable to those of SQUIDs\cite{J. C. Allred2002, I. K. Kominis2003}. Among the various types of AMs, the spin-exchange relaxation-free (SERF) magnetometer ranks as one of the most sensitive magnetometers in the world\cite{I. K. Kominis2003, D. Budker2007}. In contrast to SQUIDs, AMs can simultaneously realize miniaturization, low consumption, excellent temporal and spatial resolution, portability, and ultra-high sensitivity \cite{S. J. Seltzer2004, V. Shah2007, T. H. Sander2012}.

Miniaturization, portability, and low cost are primary development targets for bio-magnetic imaging technologies based on AMs or SQUIDs. In 2004, the National Institute of Standards and Technology (NIST) demonstrated a chip-scale AM by utilizing a microelectromechanical system approach to enable small size and low power consumption \cite{P. D. D. Schwindt2004, P. D. D. Schwindt2007}. QuSpin, Inc. presented a commercial, compact, high-performance zero-field AM (QuSpin QZFM) in 2013, and related bio-magnetic imaging tests for magnetoencephalography (MEG) and magnetocardiography (MCG) are underway\cite{V. K. Shah2013, E. Boto2017, J. Scheuer2018}. 
In bio-magnetic imaging applications, multiple miniature AM sensors are needed to form a magnetic field measurement array. The sensors are usually separate and independent, with each sensor consisting of a vertical cavity surface emitting laser (VCSEL), a vapor cell, and a photodetector (PD)\cite{P. D. D. Schwindt2004, J. Kitching2008}. When operating in array mode, differential mode noise is introduced owing to the non-uniformity of VCSEL light sources, vapor cells, and PDs, which deteriorates the imaging resolution of the magnetic field. The imaging quality tends to deteriorate more as the number of sensors increases. Moreover, the integration of a laser source and PD into a probe head not only increases the sensor size and the cost per channel, but also inevitably introduces electromagnetic noise, which increases the noise floor of an AM and reduces its sensitivity.

In this paper, we propose a novel AM design for bio-magnetic imaging systems, consisting of a double-pass optical path configuration based on a mirror and an optical fiber circulator. Without a VCSEL light source and PD integrated into the sensor head, each sensor channel shares one common laser source. When arrayed to build a multi-channel bio-magnetic imaging system, these sensors may have good uniformity of common-mode noise, which allows high sensitivity and spatial resolution for imaging bio-magnetic fields.

\section{Principle}
\label{sec:Principle}

A classical AM consists of three crucial elements: a pumping laser, an atomic ensemble (usually consisting of alkali metal), and a detection system\cite{T. M. Tierney2019}. The pumping laser optically pumps the atomic ensemble to a high steady-state polarization along the laser axis. In the presence of an external magnetic field, the polarization of the atomic ensemble is changed into a coherent spin precession. Then, the orientation or magnitude of the atomic spin polarization is measured by the detection system, and thus, the magnetic field is obtained\cite{D. Budker2007, M. P. Ledbetter2008,S. J. Seltzer2004,D. Budker2013,J. Dupont-Roc1969}. The sensitivity of an AM is limited by the lifetime of the coherent spin precession as follows \cite{J. C. Allred2002}:
\begin{equation}
	\delta B=\frac{1}{\gamma \cdot \sqrt{N \cdot \tau \cdot t}},
	\label{eq:equ1}
\end{equation}
where $\gamma$ is the gyromagnetic ratio of the alkali metal in use, $N$ is the number of atoms being polarized and detected, $\tau$ is the coherent relaxation time, and $t$ is the integration time.

When special precautions are taken to ensure a long lifetime for the atomic spin coherence, any perturbations of this coherence, such as those induced by external magnetic fields, can be detected with high sensitivity\cite{W. Happer1973,S. Appelt1999}. However, this coherent spin precession is disrupted by many factors, such as the presence of an ambient magnetic field or collisions with the wall of the vapor cell. Herein, the dominating factor is the spin-exchange process, which leads to destruction of the spin coherence phase when two atoms with different hyperfine states ($F=I\pm1/2$) collide with each other\cite{W. Happer1973,S. J. Seltzer2008,Y. Guo2019}.

When the spin-exchange collision rate exceeds the precession frequency (Larmor frequency), the relaxation due to spin exchange can be greatly suppressed, corresponding to the SERF regime demonstrated by M.V. Romalis $et\ al.$ in 2002\cite{J. C. Allred2002,H. B. Dang2010}.

In the SERF regime, we define a coordinate system with the direction of laser beam propagation along $z$ and the external magnetic field along $y$.
The behavior of the spin polarization $S$ can be described phenomenologically using the Bloch equations as\cite{S. J. Seltzer2008,S. P. Krzyzewski2019,T. M. Tierney2019}
\begin{equation}
	\frac{\mathrm{d}}{\mathrm{d} t} S=\frac{1}{q}\left[\gamma_{e} B \times S+R_{op}\left(\frac{1}{2} s \hat{z}-S\right)-R_{rel} S\right],
	\label{eq:equ2}
\end{equation}
where $B$ is the external magnetic field, $q$ is the nuclear slowing-down factor (which depends on the spin polarization), $\gamma_{e}=g_{s} \mu_{B} / \hbar$ is the electron gyromagnetic ratio, $g_{s}$ is the electron Lande factor, $\mu_{B}$ is the Bohr magneton, $R_{op}$ is the optical pumping rate, $s$ is the photon polarization along the direction of propagation of the pump with magnitude equal to the degree of circular polarization, and $R_{rel}$ is the sum of all depolarization rates except the optical pumping rate.

The optical pumping rate $R_{op}$ is defined as the average rate at which an unpolarized alkali metal atom absorbs a photon of the pumping laser, given by\cite{B. Larson1991,M. E. Wagshul1994, B. Driehuys1996}
\begin{equation}
	R_{P}=r_{e} c f_{D1} \phi s \frac{\Gamma / 2}{\left(v-v_{0}\right)^{2}+(\Gamma / 2)^{2}}
\end{equation}
where $r_{e}$ is the classical electron radius, $c$ is the velocity of light, $f_{D1}$ is the oscillation intensity of the $^{87}$Rb D1 line, $\phi$ is the photon flux, $\Gamma$ is the pressure breadth of alkali metal atoms with the quenching gas and buffer gas, $v$ is the pumping laser frequency, and $v_{0}$ is the resonance frequency of alkali metal atoms. The optical pumping rate is proportional to the power of the pumping laser and determines the atomic spin polarization with the relaxation rate and the external magnetic field simultaneously.

If the atoms are exposed to a magnetic field transverse to the laser beam, which consists of a DC component $B_{y}$ and an oscillating component $B_{mod} \cos \omega t$, the first harmonic of atomic polarization along the laser beam can be approximated as\cite{S. P. Krzyzewski2019,C. Cohen-Tannoudji1970}
\begin{equation}
	S_{z}(t)=S_{0}^{\prime} J_{0}\left(\frac{\gamma B_{mod}}{\omega}\right) J_{1}\left(\frac{\gamma B_{mod}}{\omega}\right) \frac{\gamma B_{y} \tau}{1+\left(\gamma B_{y} \tau\right)^{2}} \sin \omega t
	\label{eq:equ3}
\end{equation}
where $J_{n} (n=1,2)$ are Bessel functions of the first kind, $S_{0}^{\prime}=R_{op}/(R_{op}+R_{rel})$ is the effective polarization, $\tau=q/(R_{op}+R_{rel})$ is the coherent relaxation time, and $\gamma=\gamma_{e}/q$ is the gyromagnetic ratio. 

The power $I$ of the transmitted laser beam through the atoms is proportional to the spin polarization, given by
\begin{equation}
	\frac{d I}{d z}=-n \sigma(v) I(z)\left(1-\left\langle S_{z}\right\rangle\right)
\end{equation}
\begin{equation}
	I_{out}=I_{0} e^{-\left(1-S_{z}\right) \cdot O D}
\end{equation}
where $n$ is the atom density, $\sigma(v)$ is an absorption coefficient related to the laser frequency $v$, and $I_{out}$ and $I_{0}$ are the power of the incident and transmitted laser beams, respectively. The optical depth $O D=n \sigma(v) l$, where $l$ is the length of the laser propagating through the atomic vapor cell.

\begin{figure}[!htbp]
	
	\centerline{\includegraphics[width=3.8 in]{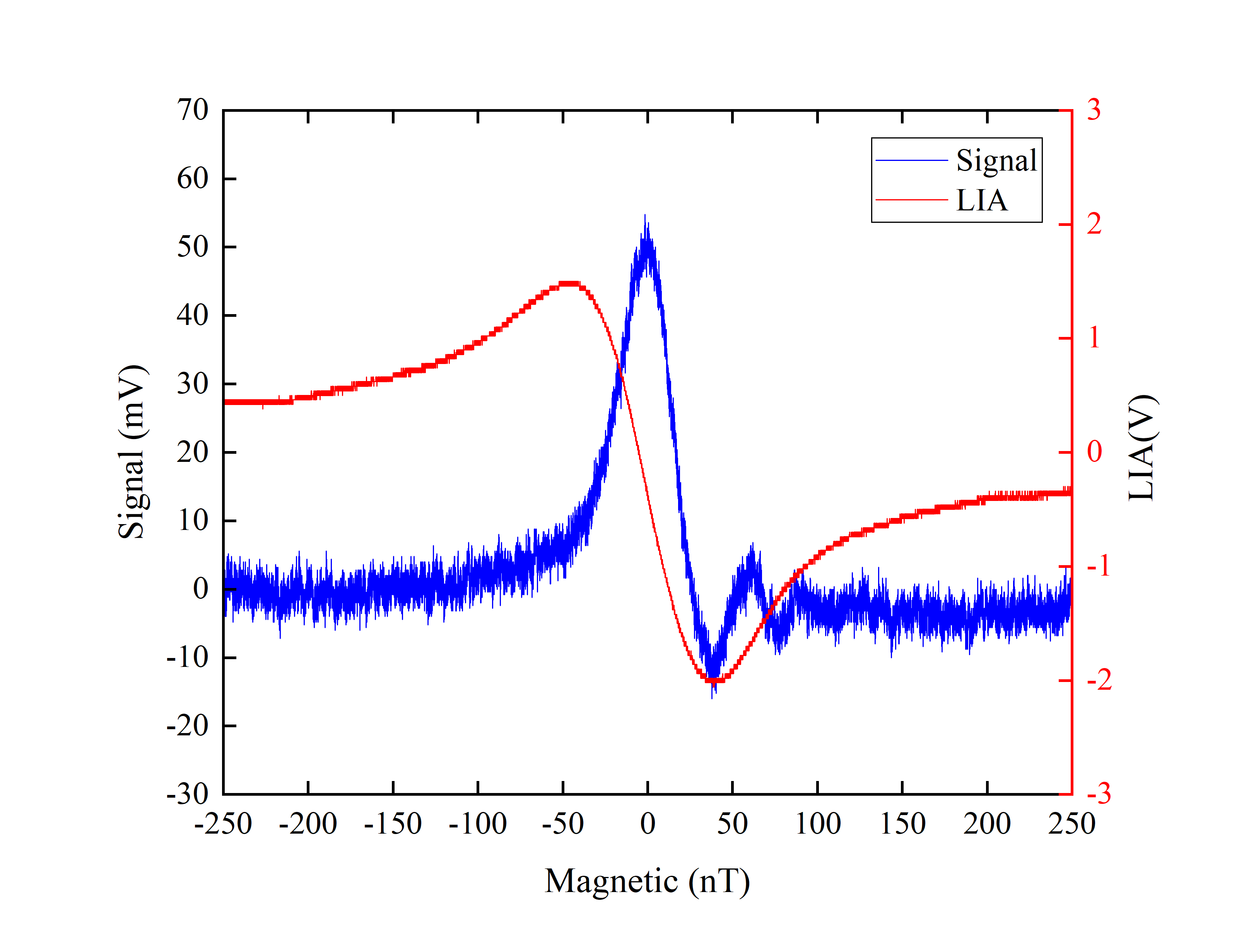}}
	\caption{A typical experimental resonance curve (blue) and corresponding error signal (red) in the zero-field SERF regime. The SERF line-width (fullwidth at half-maximum, FWHM) is approximately 34 nT for a temperature of 140 $^\circ$C and an incident laser power of 80 $\mu$W.}
	\label{Signal_LIA}
\end{figure}

Equations (2)-(6) constitute control equations of a SERF AM system with magnetic field modulation and lock-in detection. A typical experimental resonance curve is shown in Fig.~\ref{Signal_LIA}.

\section{Experimental Setup}
\label{sec:Experimental Setup}

\begin{figure*}[!htbp]
	\centerline{\includegraphics[width=7 in ]{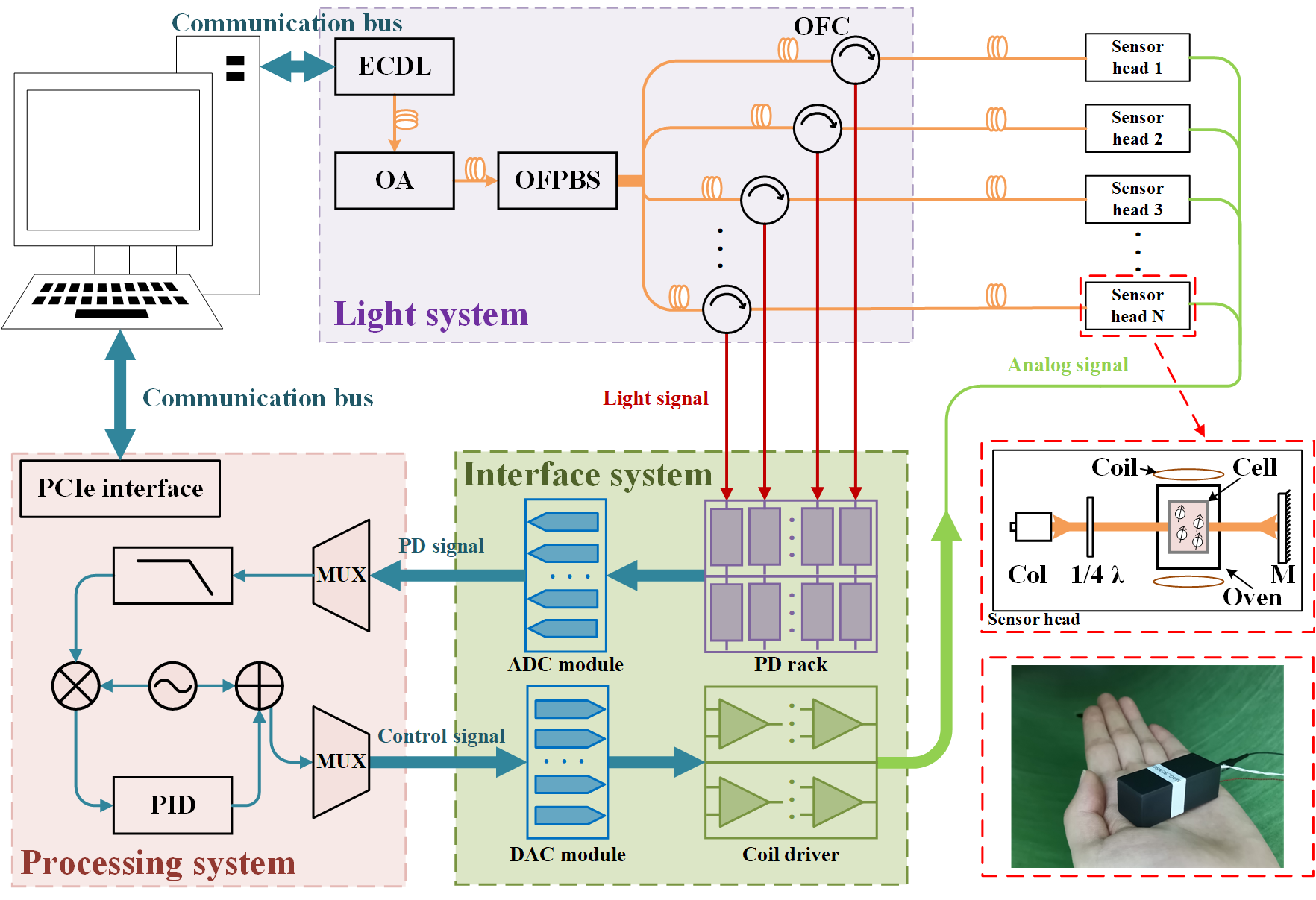}}
	\caption{Schematic of a bio-magnetic imaging system and SERF AM. ECDL: external cavity diode laser; OA: optical amplifier; OFPBS: optical fiber power beam-splitter; OFC: optical fiber circulator; Col: collimator; $\lambda$/4: quarter-wave plate; M: mirror; PD rack: photodetector rack; Mux: multiplexer. A schematic and a photograph of the sensor head are shown in the red dashed box.}
	\label{system}
\end{figure*}

The design of a bio-magnetic imaging system with SERF AMs is schematically shown in Fig.~\ref{system}.

The bio-magnetic imaging system consists of a laser source, multi-channel fiber power beam-splitter, sensor heads, PD rack, and control$\&$signal-processing unit. One 795 nm external cavity diode laser (ECDL, DL Pro, TOPTICA) is used for the laser pumping and detecting source for multiple AM sensor heads. This cavity diode laser is locked to the resonance transition of the $^{87}$Rb D1 line based on the Zeeman modulation frequency stabilization method\cite{R. A. Valenzuela1988}. After being coupled into a fiber coupler and distributed into multiple channels by the fiber power beam-splitter, the laser beam exiting out of each channel enters a sensor head via Port 1 of an optical fiber circulator. The sensor head is an all-optical configuration with a single beam\cite{V. Shah2007} and a double-pass AM design, as shown in the inset in Fig.~\ref{system}.

The input laser for the sensor head comes from Port 2 of the optical fiber circulator and is collimated with a titanium collimator (60FC-4-M10-02, Schafter+Kirchhoff). After passing through a linear polarizer, quarter-wave plate, and vapor cell, the laser is retro-reflected by a mirror, double-passed through the vapor cell, and coupled into the same collimator for detection. The detection laser is finally collected by a PD (PDA36A, THORLABS)) in the PD rack via Port 3 of the optical fiber circulator. A set of solenoid coils is placed to produce a modulating and compensating magnetic field along the direction perpendicular to the laser axis.  The vapor cell is a $4\ \rm{mm}\times 4\ \rm{mm}\times 4\ \rm{mm}$ cubic cell, containing a droplet of Rb, 100 torr of quenching gas $N_2$, and 700 torr of buffer gas $^{4}He$. The vapor cell is placed in a miniature oven fixed in the sensor head and is heated to approximately 150 $^\circ$C by two heating films with a 50 kHz AC heating current. The oven is made from polytetrafluoroethylene (PTFE) and is supported by an aluminum frame, which also holds the mirror and collimator to maintain structural stability in the retro-reflected configuration. During assembly of the sensor head, the position of the mirror is carefully adjusted by a precision fiber optic alignment stage to ensure that the laser is coupled back into the optical fiber through the collimator.

All components were carefully designed with non-magnetic materials, including wave plates, vapor atomic cells, heater films, mirrors, coils, and supporting structures.

\section{Results and Discussion}
\label{sec:Results and Discussion}

In our experiments, a zero-field resonance signal is provided by a lock-in amplifier (LIA, MFLI, Zurich Instruments). A signal generator (33522B, KEYSIGHT) provides reference signals for the modulation and LIA loop. The zero-field signal from the LIA is acquired by a data acquisition board and a closed loop locked by a digital proportional–integral–derivative controller (PID). The double-pass magnetometer sensor is placed in the center of a five-layer magnetic shield with a shielding factor larger than $10^5$, which provides a zero-field environment.

When a magnetic field orthogonal to the pumping axis is scanned in terms of amplitude near a zero magnetic field, the transparency of the atomic cell exhibits a Lorentzian curve, as shown in Fig.~\ref{Signal_LIA}. The blue Lorentzian curve measured with the magnetometer has a line-width (FWHM) of approximately 34 nT. A small modulation in the magnetic field at approximately 4 kHz is applied by the internal coils. With a phase-sensitive LIA, the PD output is demodulated to produce a dispersion curve, as shown by the red line. The amplitude and line-width of the absorption and dispersion curve determine the sensitivity of the magnetometer, with the latter being related to the relaxation rate of atoms from their pumped state via physical processes, where spin-exchange collisions provide the dominant effect. 
By heating the vapor cell, the magnetometer works in a SERF regime in which the rate of spin-exchange collisions (due to high atomic density) greatly exceeds the precession frequency (which is low during operations at very low fields), resulting in a narrow line-width resonance.

Compared with the traditional single-pass configuration\cite{V. K. Shah2013}, the present retro-reflected configuration has a longer interaction time between the laser and atoms because the laser passes through the vapor cell twice. Without the optical fiber circulator and mirror, the laser in the single-pass sensor passes through the cell once and is detected directly by a PD integrated in the sensor head. We experimentally measured and compared the performance difference between two configurations in terms of the line-width of the SERF resonance curve, considering effects of the cell temperature and laser power.

\begin{figure}[!htbp]
	\centerline{\includegraphics[width=\columnwidth]{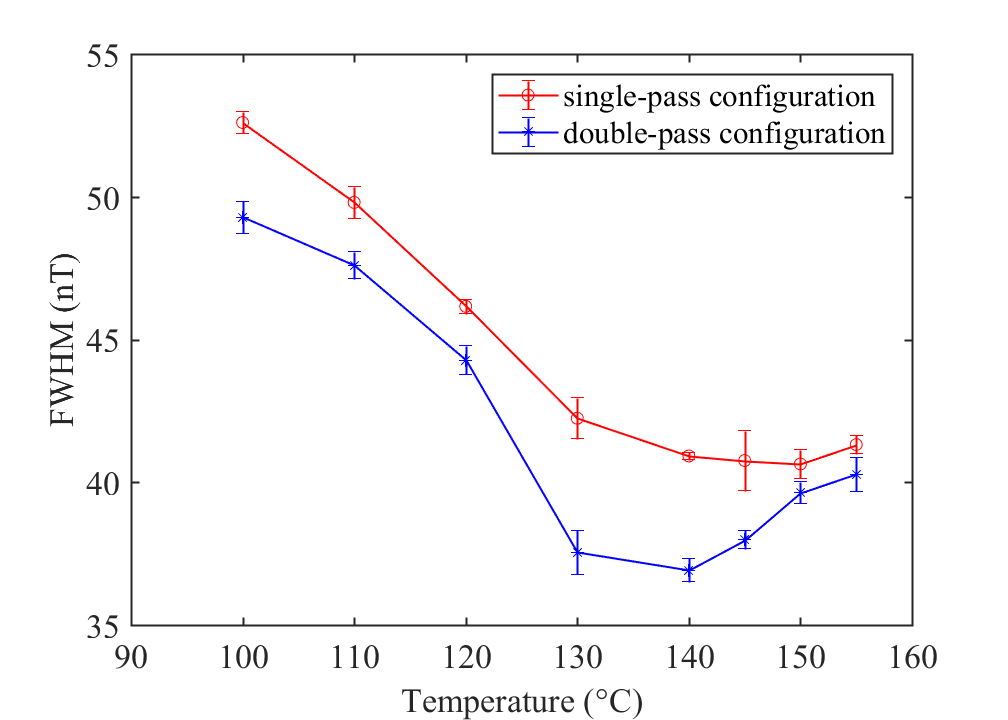}}
	\caption{Effect of cell temperature on the magnetic resonant line-width for two configurations. The power of the incident laser is maintained at 40 $\mu$W for both cases. The red and blue symbols represent experimental data for the double-pass and single-pass configurations, respectively.}
	\label{FWHM_Temp}
\end{figure}

Figure~\ref{FWHM_Temp} shows the effect of cell temperature on the magnetic resonant line-width for two configurations. The power of the incident laser is maintained at 40 $\mu$W for both cases. As the cell temperature rises above 90 $^\circ$C, spin-exchange relaxation is suppressed, and the line-width of the resonance curve decreases for both configurations. For the double-pass configuration, the resonance line-width tends to have a minimum of approximately 36 nT when the cell temperature is near 140 $^\circ$C. As the cell temperature increases above 140 $^\circ$C, the resonance line-width increases owing to the influence of spin destruction relaxation\cite{J. C. Allred2002,I. K. Kominis2003}. This finding confirms that the AM is working in the SERF regime. As shown in Fig.~\ref{FWHM_Temp}, the resonance line-width of the double-pass sensor is much narrower than that of the single-pass sensor because of the longer interaction time between the laser and the atoms. The data shown here are the average of triplicate measurements performed under the same experimental conditions. The corresponding errors are also shown in the figure.

\begin{figure}[!htbp]
	\centerline{\includegraphics[width=\columnwidth]{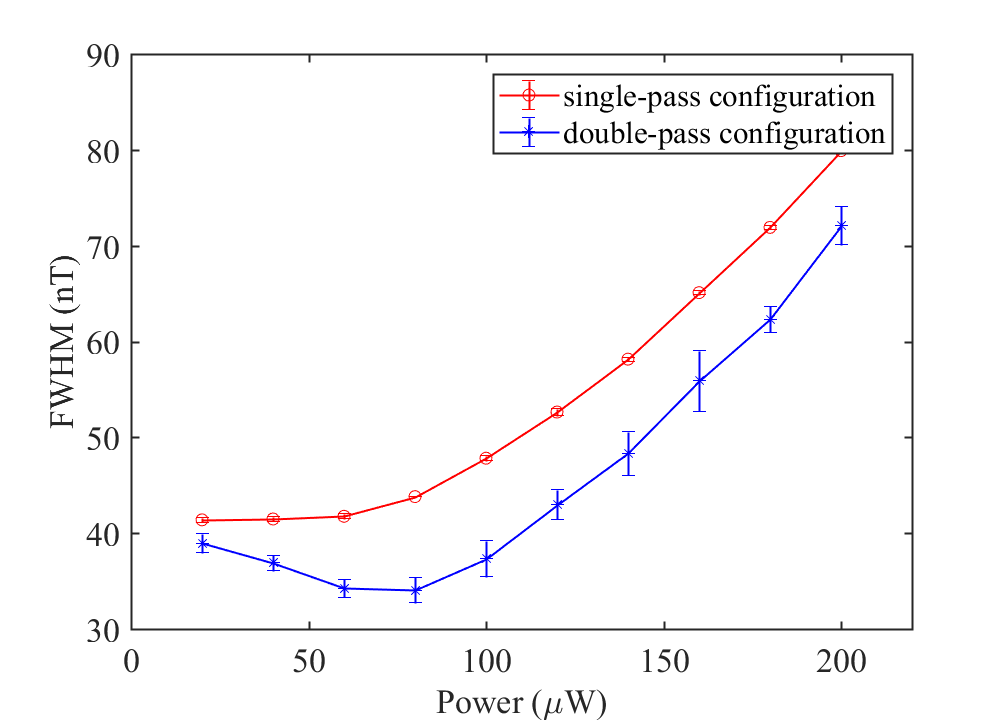}}
	\caption{Effect of pumping laser power on the magnetic resonance line-width for two configurations. The cell temperature is maintained at 140 $^\circ$C in both cases. The red and blue symbols represent experimental data for the double-pass and single-pass configurations, respectively.}
	\label{FWHM_Power}
\end{figure}

Figure~\ref{FWHM_Power} shows the effect of pumping laser power on the magnetic resonance line-width for two configurations. In the experiment, the power of the pumping laser was varied from 20 $\mu$W to 200 $\mu$W, and the line-width of the resonance curve was measured while the cell temperature was maintained at 140 $^\circ$C. The resonance line-widths tend to decrease with increasing laser power for low laser powers, reaching a minimum of approximately 34 nT at 80 $\mu$W. The line-widths then increase with laser power for both configurations. The atomic polarization is a result of the atomic relaxation rate and the optical pumping rate, which is related to the pumping laser power. When the laser power is low, the optical pumping rate is small compared with the relaxation rate, and the polarization increases as the laser power rises. Meanwhile, for a high pumping condition, most of the atoms in the cell are pumped and polarized in a saturation condition, and the atomic ensemble may not work in the SERF regime. In this case, the resonance line-width rises with increasing pumping laser power. Because the laser passes through the atomic vapor cell twice in the double-pass configuration and its pumping time and polarization efficiency are higher, the line-widths are lower than those in the single-pass configuration for a given laser power. Similar to our previous experiment, the data shown herein are the average of triplicate measurements performed under the same experimental conditions.

\begin{figure}[!htbp]
	\centerline{\includegraphics[width=\columnwidth]{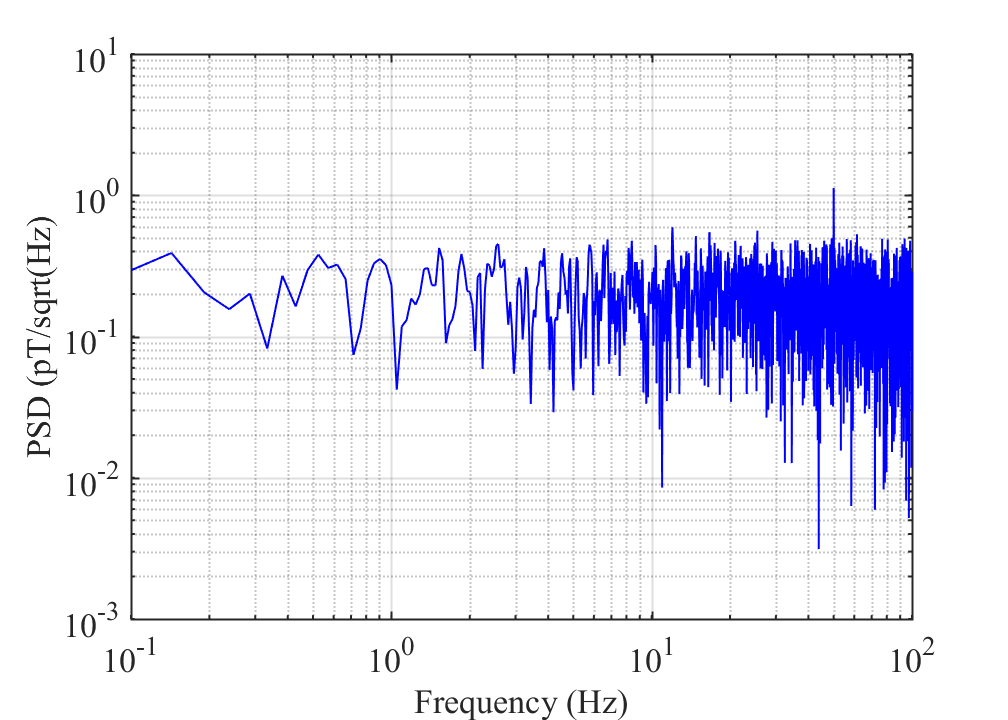}}
	\caption{Noise power spectrum density of the zero-field resonance signal. The sensitivity of this SERF AM is measured to be approximately 120 fT/Hz$^{1/2}$ at a bandwidth of 10 Hz.}
	\label{T106}
\end{figure}

As shown in Fig.~\ref{T106}, the sensitivity of the double-pass AM is characterized by its noise power spectrum when operating in a closed-loop mode with a cell temperature of 140 $^\circ$ and laser power of 80 $\mu$W. A 50 Hz peak exists in the spectrum, corresponding to the industrial frequency noise of magnetic fields. The sensitivity of this double-pass AM is measured to be approximately 120 fT/Hz$^{1/2}$ at a bandwidth of 10 Hz. This sensitivity is limited by several possible noise sources, including fluctuations of the ambient field, laser power, and temperature; vibrations of the detecting apparatus; and electrical noise in the signal-processing circuits.

\section{Conclusion}
In summary, we have proposed a new design for a bio-magnetic imaging system, based on closed-loop SERF AMs with a single-beam, double-pass configuration. A double-pass SERF sensor head with an all-optical-fiber structure has been experimentally demonstrated and characterized in comparison with the traditional single-pass configuration. The effects of cell temperature and pumping laser power on the zero-field magnetic resonance line-width were investigated. The typical resonance line-width was measured and calculated to be approximately 34 nT at a laser power of 80 $\mu$W and a cell temperature of 140 $^\circ$C. The noise power spectrum of the closed-loop SERF magnetometer shows that the sensitivity of the magnetic field reaches approximately 120 fT/Hz$^{1/2}$ at a bandwidth of 10 Hz.

The experimental results show that the double-pass configuration can achieve better performance than the traditional single-pass configuration. Additionally, the all-optical-fiber design allows for a small, compact sensor head and enables a sensor array layout with high spatial intensity, which is important for biological magnetic field measurements, such as MEG and MCG.

This design also presents a low-cost solution for bio-magnetic field imaging applications, with one laser source used for a multi-channel AM sensor array. The use of one laser source for multiple sensors is helpful for suppressing common-mode noise, for improving the sensitivity of the sensor, and for enhancing the flexibility of the bio-magnetic field image. Future work will focus on improving the sensitivity of the AM and developing an imaging system based on a multi-channel sensor array for biological applications.


\end{document}